\documentclass{JHEP3}

  \usepackage{latexsym,bm,amsmath,amssymb,amsfonts}
  \usepackage{epsfig,graphics,graphicx}
  \usepackage{slashed}
  \usepackage[latin1]{inputenc}
  \usepackage{mathrsfs}
\usepackage{mathcomp}

  \long\def\comment#1{ }

  \newcommand{\arctanh}{\mbox{arctanh}}
  \newcommand{\eqnum}[1]{Eq.~\eqref{#1}}

  \newcommand{\del}{\partial}

  \newcommand{\mcal}{\mathcal}
  \newcommand{\rme}{{\rm e}}
  \newcommand{\rmd}{{\rm d}}   

  \newcommand{\nn}{\nonumber\\}
   
  \newcommand{\order}[1]{\mcal{O}{(#1)}}

  \newcommand{\beq}{\begin{eqnarray}}
  \newcommand{\eeq}{\end{eqnarray}}
  
 \def\simge{\mathrel{%
   \rlap{\raise 0.511ex \hbox{$>$}}{\lower 0.511ex \hbox{$\sim$}}}}
\def\simle{\mathrel{
   \rlap{\raise 0.511ex \hbox{$<$}}{\lower 0.511ex \hbox{$\sim$}}}}

\vspace{1.5cm}

\title{\rm \LARGE \bf Stochastic trailing string and Langevin dynamics
from AdS/CFT}

\author{G. C. Giecold and E. Iancu\\Institut de Physique Th\'eorique,
CEA Saclay, CNRS (URA 2306),
 F-91191 Gif-sur-Yvette, France\\
  E-mail: \email{gregory.giecold@cea.fr},
        \email{edmond.iancu@cea.fr}}
\author{A. H.~Mueller\\Department of Physics, Columbia University, New York, NY
10027, USA\\
        E-mail: \email{amh@phys.columbia.edu}}

\abstract{Using the gauge/string duality, we derive a set of Langevin
equations describing the dynamics of a relativistic heavy quark moving
with constant average speed through the strongly--coupled ${\mathcal
N}\!=\!4$ SYM plasma at finite temperature. We show that the stochasticity
arises at the string world--sheet horizon, and thus is causally
disconnected from the black hole horizon in the space--time metric.
This hints at the non--thermal nature of the fluctuations, as further
supported by the fact that the noise term and the drag force
in the Langevin equations do not obey the Einstein relation.
We propose a physical picture for the dynamics of the heavy
quark in which dissipation and fluctuations are interpreted as
medium--induced radiation and the associated quantum--mechanical
fluctuations. This picture provides the right parametric estimates
for the drag force and the (longitudinal and transverse) momentum
broadening coefficients.}

\begin{document}

\section{Introduction}
\setcounter{equation}{0}

Motivated by possible strong--coupling aspects in the dynamics of
ultrarelativistic heavy ion collisions, there have been many recent
applications of the AdS/CFT correspondence to the study of the response
of a strongly coupled plasma --- typically, that of the ${\mathcal
N}\!=\!4$ supersymmetric Yang--Mills (SYM) theory at finite temperature
--- to an external perturbation, so like a ``hard probe'' --- say, a heavy
quark, or an electromagnetic current (see the review papers
\cite{Son:2007vk,Iancu:2008sp,Gubser:2009sn} for details and more
references). Most of these studies focused on the mean field dynamics
responsible for dissipation (viscosity, energy loss, structure
functions), as encoded in retarded response functions --- typically, the
2--point Green's function of the ${\mathcal N}\!=\!4$ SYM operator
perturbing the plasma. By comparison, the statistical properties of the
plasma (in or near thermal equilibrium) have been less investigated.
Within the AdS/CFT framework, such investigations would require field
quantization in a curved space--time ---  the $AdS_5\times S^5$
Schwarzschild geometry dual to the strongly--coupled ${\mathcal N}\!=\!4$
SYM plasma ---, which in general is a very difficult problem. Still,
there has been some interesting progress in that sense, which refers to a
comparatively simpler problem: that of the quantization of the small
fluctuations of the Nambu--Goto string dual to a heavy quark immersed
into the plasma.

Several noticeable steps may be associated with this progress: In Ref.
\cite{Herzog:2002pc}, a prescription was formulated for computing the
Schwinger--Keldysh Green's functions at finite temperature within the
AdS/CFT correspondence. With this prescription, the quantum thermal
distributions are generated via analytic continuation across the horizon
singularities in the Kruskal diagram for the $AdS_5$ Schwarzschild
space--time. Using this prescription, one has computed the diffusion
coefficient of a non--relativistic heavy quark
\cite{CasalderreySolana:2006rq}, and the momentum broadening for a
relativistic heavy quark which propagates through the plasma at constant
(average) speed \cite{CasalderreySolana:2007qw,Gubser:2006nz}. Very
recently, in Refs.~\cite{deBoer:2008gu,Son:2009vu}, a set of Langevin
equations has been constructed which describes the Brownian motion of a
non--relativistic heavy quark and of the attached Nambu--Goto string.
Within these constructions, the origin of the `noise' (the random force
in the Langevin equations) in the supergravity calculations lies at the
black hole horizon, as expected for thermal fluctuations.

The Langevin equations in Refs.~\cite{deBoer:2008gu,Son:2009vu} encompass
previous results for the drag force \cite{Herzog:2006gh,Gubser:2006bz}
and the diffusion coefficient \cite{CasalderreySolana:2006rq} of a {\em
non--relativistic} heavy quark. But to our knowledge, no attempt has been
made so far at deriving corresponding equations for a {\em relativistic}
heavy quark, whose dual description is a trailing string
\cite{Herzog:2006gh,Gubser:2006bz}. In fact, the suitability of the
Langevin description for the stochastic trailing string was even
challenged by the observation that the respective expressions for the
drag force and the momentum broadening do not to obey the Einstein
relation \cite{Gubser:2006nz}. The latter is a hallmark of thermal
equilibrium and must be satisfied by any Langevin equation describing
thermalization. However, Langevin dynamics is more general than
thermalization, and as a matter of facts it does apply to the stochastic
trailing string, as we will demonstrate in this paper.

Specifically, our objective in what follows is twofold: \texttt{(i)} to
show how the Langevin description of the stochastic trailing string
unambiguously emerges from the underlying AdS/CFT formalism, and
\texttt{(ii)} to clarify the physical interpretation of the associated
noise term, in particular, its non--thermal nature.

Our main conclusion is that the stochastic dynamics of the relativistic
quark is fundamentally different from the Brownian motion of a
non--relativistic quark subjected to a thermal noise. Within the
supergravity calculation, this difference manifests itself via the
emergence of an event horizon on the string world--sheet
\cite{CasalderreySolana:2007qw,Gubser:2006nz}, which lies in between the
Minkowski boundary and the black hole horizon, and which governs the
stochastic dynamics of the fast moving quark. With our choice for the
radial coordinate $z$ in $AdS_5$, the Minkowski boundary lies at $z=0$,
the black hole horizon at $z_H=1/T$, and the world--sheet horizon at
$z_s=z_H/\sqrt{\gamma}$, where $\gamma=1/\sqrt{1-v^2}$ is the Lorentz
factor of the heavy quark. (We assume that the quark is pulled by an
external force in such a way that its average velocity remains constant.)
The presence of the world--sheet horizon means that the dynamics of the
upper part of the string at $z<z_s$ (including the heavy quark at
$z\simeq 0$) is causally disconnected from that of its lower part at
$z_s<z<z_H$, and thus cannot be influenced by thermal fluctuations
originating at the black hole horizon.

This conclusion is supported by the previous calculations of the momentum
broadening for the heavy quark
\cite{CasalderreySolana:2007qw,Gubser:2006nz}, which show that the
relevant correlations are generated (via analytic continuation in the
Kruskal plane) at the {\em world--sheet} horizon, and not at the black
hole one. Formally, these correlations look as being thermal (they
involve the Bose--Einstein distribution), but with an effective
temperature $T_{\rm eff}=T/\sqrt{\gamma}$, which is the Hawking
temperature associated to the world--sheet horizon. Thus, no
surprisingly, our explicit construction of the Langevin equations will
reveal that the corresponding noise terms arise from this world--sheet
horizon.

The Langevin equations for the relativistic heavy quark will be
constructed in two different ways: \texttt{(1)} by integrating out the
quantum fluctuations of the upper part of the string, from the
world--sheet horizon up to the boundary, and \texttt{(2)} by integrating
out the string fluctuations only within an infinitesimal strip in $z$,
from the world--sheet horizon at $z=z_s$ up to the `stretched' horizon at
$z=z_s(1-\epsilon)$ with $\epsilon\ll 1$; this generates a `bulk' noise
term at the stretched horizon, whose effects then propagate upwards the
string, via the corresponding classical solutions. Both procedures
provide exactly the same set of Langevin equations, which encompass the
previous results for the drag force \cite{Herzog:2006gh,Gubser:2006bz}
and for the (longitudinal and transverse) momentum broadening
\cite{CasalderreySolana:2007qw,Gubser:2006nz}. In these manipulations,
the lower part of the string at $z>z_s$ and, in particular, the black
hole horizon, do not play any role, as expected from the previous
argument on causality.

If the relevant fluctuations are not of thermal nature, then why do they
{\em look} as being thermal ? What is their actual physical origin ? And
what is the role played by the thermal bath ? To try and answer such
questions, we will rely on a physical picture for the interactions
between an energetic parton and the strongly--coupled plasma which was
proposed in
Refs.~\cite{Hatta:2007cs,Hatta:2008tx,Dominguez:2008vd,Iancu:2008sp}, and
that we shall here more specifically develop for the problem at hand. In
this picture, both the energy loss (`drag force') and the momentum
broadening (`noise term') are due to medium--induced radiation. This is
reminiscent of the mechanism of energy loss of a heavy, or light, quark
at weak coupling
\cite{BDMPS,Baier:2002tc,Kovner:2003zj,CasalderreySolana:2007zz}, with
the main difference being in the cause of the medium--induced radiation.
At weak coupling, multiple scattering off the plasma constituents frees
gluonic fluctuations in the quark wavefunction, while at strong coupling
the plasma exerts a force, proportional to $T^2$, acting to free quanta
from the heavy quark as radiation. In the gravity description, this
appears as a force pulling energy in the trailing string towards the
horizon. At either weak or strong coupling, quanta are freed when their
virtuality is smaller than a critical value, the {\em saturation
momentum} $Q_s$; at strong coupling and for a fast moving quark, this
scales like $Q_s\sim \sqrt{\gamma}T$. Within this picture, the
world--sheet horizon at $z_s\sim 1/Q_s$ corresponds to the causal
separation between the highly virtual quanta ($Q\gg Q_s$), which cannot
decay into the plasma and thus are a part of the heavy quark
wavefunction, and the low virtuality ones, with $Q\lesssim Q_s$, which
have already been freed, thus causing energy loss. The recoil of the
heavy quark due to the random emission of quanta with $Q\lesssim Q_s$ is
then responsible for its momentum broadening.

From his perspective, the noise terms in the Langevin equations for the
fast moving quark reflect quantum fluctuations in the emission process.
Of course, the presence of the surrounding plasma is essential for this
emission to be possible in the first place (a heavy quark moving at
constant speed through the vacuum could not radiate), but the plasma acts
merely as a background field, which acts towards reducing the virtuality
of the emitted quanta and thus allows them to decay. The genuine thermal
fluctuations on the plasma are unimportant when $\gamma\gg 1$, although
when $\gamma\simeq 1$ they are certainly the main source of
stochasticity, as shown in \cite{deBoer:2008gu,Son:2009vu}. Besides, we
see no role for Hawking radiation of supergravity quanta at any value of
$\gamma$.

This picture is further corroborated by the study of a different physical
problem, where the thermal effects are obviously absent, yet the
mathematical treatment within AdS/CFT is very similar to that for the
problem at hand: this is the problem of a heavy quark propagating with
constant acceleration $a$ through the vacuum of the strongly--coupled
${\mathcal N}\!=\!4$ SYM theory
\cite{Dominguez:2008vd,Xiao:2008nr,Paredes:2008cr}. The accelerated
particle can radiate, and this radiation manifests itself through the
emergence of a world--sheet horizon, leading to dissipation and momentum
broadening. The fluctuations generated at this horizon are once again
thermally distributed, with an effective temperature $T_{\rm
eff}=a/2\pi$. In that context, it is natural to interpret the induced
horizon as the AdS dual of the Unruh effect \cite{Unruheffect} : the
accelerated observer perceives the Minkowski vacuum as a thermal state
with temperature $a/2\pi$. For an inertial observer, this is interpreted
as follows \cite{Unruh:1983ms}: the accelerated particle can radiate and
the correlations induced by the backreaction to this radiation are such
that the excited states of the emitted particle are populated according
to a thermal distribution. Most likely, a similar interpretation holds
also for the thermal--like correlations generated at the world--sheet
horizon in the problem at hand --- that of a relativistic quark
propagating at constant speed through a thermal bath. It would be
interesting to identify similar features in other problems which exhibit
accelerated motion, or medium--induced radiation, or both, so like the
rotating string problem considered in Ref.~\cite{Fadafan:2008bq}.

The paper is organized as follows: In Sect. 2 we construct the Langevin
equations describing the stochastic dynamics of the string endpoint on
the boundary of $AdS_5$, i.e., of the relativistic heavy quark. Our key
observation is that, in the Kruskal--Keldysh quantization of the small
fluctuations of the trailing string, the stochasticity is generated
exclusively at the world--sheet horizon. This conclusion is further
substantiated by the analysis in Sect. 3 where we follow the progression
of the fluctuations along the string, from the world--sheet horizon up to
the string endpoint on the boundary. We thus demonstrate that the noise
correlations are faithfully transmitted from the stretched horizon to the
heavy quark, via the fluctuations of the string. Finally, Sect. 4
contains our physical discussion. First, in Sect. 4.1, we argue that the
Langevin equations do not describe thermalization, although they do
generate thermal--like momentum distributions, but at a fictitious
temperature which is not the same as the temperature of the plasma, and
is moreover different for longitudinal and transverse fluctuations. Then,
in Sect. 4.2, we develop our physical picture for medium--induced
radiation and parton branching, which emphasizes the quantum--mechanical
nature of the stochasticity.

\section{Boundary picture of the stochastic motion}
\setcounter{equation}{0}

In this section we will construct a set of Langevin equations for the
stochastic dynamics of a relativistic heavy quark which propagates with
uniform average velocity through a strongly--coupled ${\mathcal N}\!=\!4$
SYM plasma at temperature $T$. To that aim, we will follow the general
strategy in Ref. \cite{Son:2009vu}, that we will extend to a fast moving
quark and the associated trailing string. In this procedure, we will also
rely on previous results in the literature
\cite{CasalderreySolana:2007qw,Gubser:2006nz} concerning the classical
solutions for the fluctuations of the trailing string and their
quantization via analytic continuation in the Kruskal plane.


\subsection{The trailing string and its small fluctuations}

The AdS dual of the heavy quark is a string hanging down in the radial
direction of $AdS_5$, with an endpoint (representing the heavy quark)
attached to a D7--brane whose radial coordinate fixes the bare mass of
the quark. The string dynamics is encoded in the Nambu--Goto action,
 \beq\label{SNG}
 S\,=\,-\frac{1}{2\pi\ell_s^2}\int\rmd^2\sigma\sqrt{-\mbox{det}\,
 h_{\alpha\beta}}\,,\qquad
 h_{\alpha\beta}\,=\,g_{\mu\nu}\partial_\alpha x^\mu
 \partial_\beta x^\nu\,,\eeq
where $\sigma^\alpha$, $\alpha=1,\,2$, are coordinates on the string
world--sheet, $h_{\alpha\beta}$ is the induced world--sheet metric, and
$g_{\mu\nu}$ is the metric of the $AdS_5$--Schwarzschild space--time,
chosen as
 \beq\label{Metric}
\rmd s^{2} = \frac{R^{2}}{z_H^{2} z^{2}}\left(-f(z) \rmd t^{2}+\rmd {\bm
x}^{2}+\frac{\rmd z^{2}}{f(z)}\right),
\eeq where $f(z) = 1 -  z^{4}$ and $T = {1}/{\pi z_H}$ is the Hawking
temperature. (As compared to the Introduction, we have switched to a
dimensionless radial coordinate.)

The quark is moving along the longitudinal axis $x^3$ with constant
(average) velocity $v$ in the plasma rest frame. For this to be possible,
the quark must be subjected to some external force, which compensates for
the energy loss towards the plasma. The profile of the string
corresponding to this steady (average) motion is known as the `trailing
string'. This is obtained by solving the equations of motion derived from
\eqnum{SNG} with appropriate boundary conditions, and reads
\cite{Herzog:2006gh,Gubser:2006bz}
 \beq\label{Trail}
x_0^{3}\, =\, v t + \, \frac{v z_H}{2}\, \big(\arctan z - \arctanh z
 \big)\,.
 \eeq
In what follows we shall be interested in small fluctuations around this
steady solution, which can be either longitudinal or transverse: $x^{3} =
x_0^{3} + \delta x_\ell (t,z)$ and $x_\perp = \delta x_\perp (t,z)$. To
quadratic order in the fluctuations, the Nambu--Goto action is then
expanded as (in the static gauge $\sigma^\alpha=(t,z)$)
\beq\label{NG}
S = - \frac{\sqrt{\lambda} T z_s^{2}}{2} \int \rmd t \rmd z
\frac{1}{z^{2}} &+& \int\rmd t \rmd z P^{\alpha} \partial_\alpha \delta
x_\ell \nn &-&
 \frac{1}{2} \int \rmd t \rmd z \left[ T_{\ell}^{\alpha
 \beta}\partial_\alpha \delta x_\ell\,
 \partial_\beta \delta x_\ell + T_\perp^{\alpha \beta} \partial_\alpha
 \delta x_\perp^{i}
 \partial_\beta \delta x_\perp^{i} \right],
 \eeq
where $z_s \equiv \sqrt[4]{1 - v^{2}} = {1}/{\sqrt{\gamma}}$
and\footnote{In \eqnum{T pert}, we have corrected an overall sign error
in Eq.~(22) of Ref. \cite{Gubser:2006nz}.} \cite{Gubser:2006nz}
\beq\label{P pert}
P^{\alpha} = \frac{\pi v \sqrt{\lambda} T^{2} }{2 z_s^{2} } \left(
\begin{array}{ccc} \frac{z_H}{ z^{2} (1 - z^{4}) } \\ 1 \end{array}
\right), \eeq

\beq\label{T pert}
T_\perp^{\alpha \beta} = z_s^{4} T_{\ell}^{\alpha \beta} = -\frac{\pi
\sqrt{\lambda} T^{2}}{2 z_s^{2}} \left(
\begin{array}{ccc}
\frac{z_H}{z^{2}} \frac{1 - (zz_s)^{4}}{(1-z^{4})^{2}} & \frac{v^{2}}{1-z^{4}}
\\
\frac{v^{2}}{1-z^{4}} & \frac{z^{4} - z_s^{4}}{z_H z^{2}} \\
 \end{array} \right). \eeq
The quantities $T_{\ell,\,\perp}^{\alpha \beta}$ have the meaning of
local stress tensors on the string. At high energy, the components
$T_{\ell}^{\alpha \beta}$ of the longitudinal stress tensor are
parametrically larger, by a factor $\gamma^2\gg 1$, than the
corresponding components $T_\perp^{\alpha\beta}$ of the transverse stress
tensor. This difference reflects the strong energy--dependence of the
gravitational interactions.

Using $\partial_\alpha P^{\alpha}=0$, one sees that the term linear in
the fluctuations in \eqnum{NG} does not affect the equations of motion,
which therefore read
 \beq\label{EOM}
\partial_\alpha (T ^{\alpha \beta} \partial_\beta \psi) = 0,
 \qquad \psi = \delta x_\ell,\ \delta x_\perp, \eeq
in compact notations which treat on the same footing the longitudinal and
transverse fluctuations. Upon expanding in Fourier modes,
\beq\label{FT}
\psi (t,z) = \int_{-\infty}^{\infty} \frac{\rmd\omega}{2\pi}
\,\psi(\omega,z)\, \rme^{-i \omega t}, \eeq this yields
\beq\label{FT EOM}
\left\{ a(z) \partial^{2}_z - 2 b(\omega, z) \partial_z + c(\omega,
z)\right\} \psi (\omega, z) = 0, \eeq where
\beq\label{PolyEOM}
a(z)& =&  z(1-z^{4})^{2}(z_s^{4} - z^{4}),\nn b(\omega, z) &=& (1-z^{4})
\left[ 1 - z^{8} - v^{2}(1 - z^{2} + i \omega z_H z^{3}) \right],\nn
c(\omega, z)& =& \omega z_H z \left[ \omega z_H (1 - z^{4}) + v^{2} z^{4}
 (\omega z_H + 4 i z) \right]. \eeq
The zeroes of $a(z)$ determine the regular singular points of this
equation. In particular, the special role played by the point $z_s$ as a
world--sheet horizon becomes manifest at this level: for $z = z_s$ the
value of $\partial_z \psi (\omega, z_s)$ is determined from the equation
of motion. This means that fluctuations of the string at $z < z_s$ are
causally disconnected from those below the location of the world--sheet
horizon.

\subsection{Keldysh Green function in AdS/CFT}

\begin{figure}
\centerline{
\includegraphics[width=0.6\textwidth]{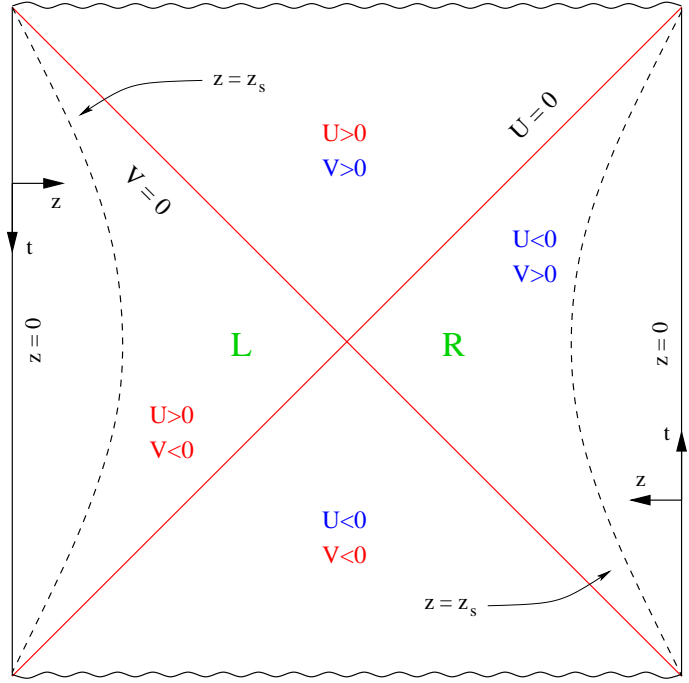}
}
\caption{\sl Kruskal diagram for $AdS_5$ Schwarzschild metric;
the position of the induced horizon on the string world--sheet is shown
with dashed lines in both $R$ and $L$ quadrants.
\label{fig:Krskl}}\end{figure}

In what follows we construct solutions to Eqs.~\eqref{FT
EOM}--\eqref{PolyEOM} for the string fluctuations which are well defined
everywhere in the Kruskal diagram for the $AdS_5$ Schwarzschild
space--time (see Fig.~\ref{fig:Krskl}). These solutions are uniquely
determined by their boundary conditions at the two Minkowski boundaries
--- in the right ($R$) and, respectively, left ($L$) quadrants of the
Kruskal diagram ---, together with the appropriate conditions of
analyticity in the Kruskal variables $U$ and $V$ (as explained in
\cite{Herzog:2002pc}). The latter amount to quantization prescriptions
which impose infalling conditions on the positive--frequency modes and
outgoing conditions on the negative--frequency ones. These prescriptions
ultimately generate the quantum Green's functions at finite--temperature
and in real time, which are time--ordered along the Keldysh contour
\cite{Herzog:2002pc}. Specifically, the time variables on the $R$ and,
respectively, $L$ boundary in Fig.~\ref{fig:Krskl} correspond to the
chronological and, respectively, antichronological branches of the
Keldysh time contour.

As usual in the framework of AdS/CFT, we are interested in the classical
action expressed as a functional of the fields on the boundary. We denote
by $\psi_R(t_R,z)$ and $\psi_L(t_R,z)$ the classical solutions in the $R$
and $L$ quadrant, respectively. Making use of the equations of motion and
integrating by parts, the classical action reduces to its value on the
boundary of the Kruskal diagram, i.e., the $R$ and $L$ Minkowski
boundaries:
\beq\label{bndry NG}
S_{\rm bndry}=
 \int \rmd t_R \Big[\!-
 P^z\psi_R  + \frac{1}{2}\,\psi_R T^{z\beta} \partial_\beta
  \psi_R\Big]_{z_R= z_m}-
   \int \rmd t_L \Big[\!-
   P^z\psi_L  +\frac{1}{2}\,\psi_L T^{z\beta} \partial_\beta
  \psi_L\Big]_{z_L= z_m}\,,\nn
 \eeq
where it is understood that the terms involving $P_z$ exist only in the
longitudinal sector. $z_m \ll 1$ is the radial location of the D7--brane
on which the string ends.

In \eqnum{bndry NG}, the world--sheet index $\beta$ can take {\em a
priori} both values $t$ and $z$, but the contribution corresponding to
$\beta=t$ is in fact zero, since the respective integrand is an odd
function of $t$. This is worth noticing since in
Ref.~\cite{Gubser:2006nz} it was found that the dominant contribution to
the imaginary part of the retarded propagator at high energy ($\gamma\gg
1$) comes from the piece proportional to $T^{zt}$. We will later see how
that contribution arises in the present calculation, where only the piece
proportional to $T^{zz}$ survives in \eqnum{bndry NG}.

Switching to the frequency representation, we introduce a basis of
retarded and advanced solutions, $\psi_{ret}(\omega, z)$ and
$\psi_{adv}(\omega, z)$, which are normalized such that $\psi_{ret}
(\omega, 0) =\psi_{adv} (\omega, 0) =1$. They obey $\psi_{ret} (\omega,
z) = \psi^{*}_{ret} (- \omega, z)$, and similarly for $\psi_{adv}$. These
solutions are truly boundary--to--bulk propagators in Fourier space. They
have been constructed in Ref.~\cite{Gubser:2006nz} (see also
\cite{CasalderreySolana:2007qw}) from which we quote the relevant
results.

Note first that, unlike what happens for a static (or non--relativistic)
quark \cite{Son:2009vu}, the retarded and advanced solutions are not
simply related to each other by complex conjugation: one rather has
 \beq
\psi_{adv} (\omega,z)\,=\,[g(z)]^{i \omega/2}\, [g({z}/{z_s})]^{-i \omega
z_H/2 z_s}\, \psi^{*}_{ret}(\omega, z)\,,\eeq
with $g(z)=\frac{1 + z}{1 - z}\, \rme^{-2 \arctan z}$. Near the boundary
($z\ll 1$), these solutions behave as follows
\beq\label{ret exp}
 \psi_{ret}(\omega, z) = \left
 (1 + \frac{z_H^{2} \omega^{2}}{2 z_s^{4}} z^{2} + \order{z^4}\right)
  + C_{ret}(\omega)\big(z^{3} + \order{z^5}\big) \eeq
\beq\label{adv exp}
 \psi_{adv}(\omega, z) = \left(1 + \frac{z_H^{2} \omega^{2}}
 {2 z_s^{4}} z^{2} + \order{z^4}\right) +
 C_{adv}(\omega)\big(z^{3} + \order{z^5}\big), \eeq
where the expansion involving even (odd) powers of $z$ is that of the
non--normalizable (normalizable) mode. The coefficients of the
normalizable mode are related by
\beq\label{Cadv}
 C_{adv}(\omega) = C^{*}_{ret}(\omega) - i X(\omega),\eeq
with
 \beq\label{X eq}
 X(\omega) \,=\,
 \frac{2\omega z_H}{3} \,v^{2}\gamma^2  \,,
 \qquad \mbox{Im}\,C_{ret}(\omega)\,=\,i\,\frac{\omega z_H}{3}\,. \eeq
The real part of coefficient $C_{ret}(\omega)$ has been numerically
evaluated in Ref.~\cite{Gubser:2006nz}. Here, we only need to know that,
at small frequency $\omega\ll z_sT$, its real part is comparatively
smaller: $\mbox{Re}\,C_{ret}(\omega)\sim\order{\omega^2z_H^2/z_s^2}$.
Note that, at high energy ($\gamma\gg 1$), $X$ dominates over
$\mbox{Im}\,C_{ret}$ in \eqnum{Cadv}.

Consider also the approach towards the world--sheet horizon ($z=z_s$)
from the above ($z<z_s$): there, $\psi_{ret}$ remains regular, whereas $
\psi_{adv}$ has a branching point:
 \beq\label{adv hor}
 \psi_{adv}(\omega, z)\,\propto\,(z_s-z)^{\frac{i \omega z_H}{2
 z_s}}\left[1+\order{z_s-z}\right]\,.\eeq
This shows that $\Psi_{adv} (\omega, t, z) \equiv \rme^{-i \omega
t}\psi_{adv}(\omega, z)$ is an outgoing wave: with increasing time, the
phase remains constant while departing from the horizon. One can
similarly argue that $\Psi_{ret} (\omega, t, z) = \rme^{-i \omega
t}\psi_{ret}(\omega, z)$ is an infalling solution \cite{Gubser:2006nz}.

We now expand the general solution in the right and left quadrants of the
Kruskal diagram in this retarded/advanced basis :
\beq\label{psi R}
\psi_R(\omega, z) &=& A(\omega) \psi_{ret}(\omega, z) + B(\omega)
\psi_{adv}(\omega, z), \nn \psi_L(\omega, z) & = &C(\omega)
\psi_{ret}(\omega, z) + D(\omega)
 \psi_{adv}(\omega, z). \eeq
We need four conditions to determine the four unknown coefficients $A$,
$B$, $C$, and $D$. Two of them are provided by the boundary values at the
$R$ and $L$ Minkowski boundaries, that we denote as
$\psi^{0}_{R}(\omega)$ and $\psi^{0}_{L}(\omega)$, respectively. The
other two are determined by analyticity conditions in the Kruskal plane,
which allows one to connect the solution in the $L$ quadrant to that in
the $R$ quadrant. With reference to Fig.~\ref{fig:Krskl}, one sees that
this requires crossing two types of horizons: the world--sheet horizons
in both $R$ and $L$ quadrants, and the black hole horizons at $U=0$ and
$V=0$. The detailed matching at these horizons, following the
prescription of Ref. \cite{Herzog:2002pc}, has been performed in Appendix
B of Ref.~\cite{Gubser:2006nz}, with the following results: the
multiplicative factors associated with crossing the black hole horizons
precisely compensate each other, unlike those associated with crossing
the world--sheet horizons, which rather enhance each other. Hence, the
net result comes from the world--sheet horizons alone\footnote{This point
is even more explicit in the analysis in
Ref.~\cite{CasalderreySolana:2007qw}, where a different set of
coordinates was used, in which the world--sheet metric is diagonal. With
those coordinates, the only horizons to be crossed when going from the
$R$ to the $L$ boundary in the respective Kruskal diagram are the {\em
world--sheet} horizons.}, and reads
 \beq\label{Modes R to L}
\left( \begin{array}{ccc} C \\ D \end{array} \right) = \left(
\begin{array}{ccc}
1 & 0 \\
0 & \rme^{\frac{\omega}{z_s T}}
\end{array} \right) \left( \begin{array}{ccc}
 A\\ B \end{array} \right) \eeq
The two independent coefficients can now be determined from the boundary
values $\psi^{0}_{R}(\omega)$ and $\psi^{0}_{L}(\omega)$. This eventually
yields
\beq\label{A bndry}
A(\omega) = (1+n)(\omega) \psi^{0}_{R}(\omega) -  n(\omega)
\psi^{0}_{L}(\omega), \eeq
\beq\label{B bndry}
B(\omega) = n(\omega) \left[ \psi^{0}_{L}(\omega) -
 \psi^{0}_{R}(\omega)\right]. \eeq
Here $n(\omega)=1/(\rme^{\omega/{z_s T}}-1)$ is the Bose--Einstein
thermal distribution with the effective temperature $T_{\rm eff} = z_s
T=T/\sqrt{\gamma}$. As it should be clear from the previous
manipulations, this effective thermal distribution has been generated via
the matching conditions at the $R$ and $L$ world--sheet horizons, cf.
\eqnum{Modes R to L}.

The equations simplify if one introduces `average' (or `classical') and
`fluctuating' variables, according to $\psi_r \equiv (\psi_R +
\psi_L)/{2}$ and $\psi_a\equiv \psi_R - \psi_L$, and similarly for the
boundary values. One then finds
 \beq\label{psi r}
\psi_{r}(\omega, z) & =& \psi^{0}_{r}(\omega) \psi_{ret}(\omega, z) +
\frac{1+2n(\omega)}{2}\, \psi^{0}_{a}(\omega) (\psi_{ret}(\omega, z) -
\psi_{adv}(\omega, z)),\nn
 \psi_{a}(\omega, z) &=& \psi^{0}_{a}(\omega)
\psi_{adv}(\omega, z),
 \eeq
and the boundary action takes a particularly simple form:
 \beq\label{G NG0}
S_{\rm bndry} & = & \frac{1}{2} \int\frac{\rmd\omega}{2 \pi}\, T^{zz}(z)
\big[ \psi_r (-\omega, z) \partial_z \psi_a (\omega, z) +
 (r\leftrightarrow a) \big]_{z= z_m}. \eeq
(We have here omitted the term linear in the fluctuations, since this
does not matter for the calculation of the 2--point Green's functions.
This term will be reinserted in the next subsection.) By combining the
above equations, we finally deduce
\beq\label{G NG}
i S_{\rm bndry} = - i \int \frac{\rmd\omega}{2 \pi}\, \psi^{0}_a (-
\omega) G^{0}_R (\omega)  \psi^{0}_r (\omega) - \frac{1}{2} \int
\frac{\rmd\omega}{2 \pi}\, \psi^{0}_a (- \omega) G_{\rm sym}(\omega)
\psi^{0}_a (\omega),
 \eeq
with the retarded and symmetric Green's functions defined as
\beq\label{G_R}
G^{0}_{\perp,\, R}(\omega) = z_s^{4} G^{0}_{\ell,\, R}(\omega) \,=\, -
\frac{1}{2}\, T_{\perp}^{zz}(z)\,\partial_z\Big[ \psi_{ret}(\omega,z)
 \psi_{adv}(-\omega,z)\Big]_{z= z_m}, \eeq
and, respectively,
 \beq\label{G_sym}
 G_{\rm sym}(\omega) = - (1 + 2 n(\omega)) \,\mbox{Im}\,G_R^0(\omega). \eeq
Note that \eqnum{G_sym} is formally the same as the
fluctuation--dissipation theorem (or `KMS condition') characteristic of
thermal equilibrium, but with an effective temperature $T_{\rm eff} = z_s
T$. By also using Eqs.~(\ref{ret exp})--(\ref{adv exp}) together with the
expression of $T_{\perp}^{zz}$ given in \eqref{T pert}, one finally
deduces
\beq\label{GRplicit}
G^{0}_{\perp,\,R}(\omega)\,=\, z_s^{4} G^{0}_{\ell,\, R} \,=
\,G_{R}(\omega) - \gamma
 M_{Q}\omega^{2}\,,\eeq
where
 \beq\label{GRfin}
 G_{R}(\omega)&\equiv &
 -Y \big(C_{ret}(\omega)+C_{adv}^*(\omega)\big),\nn
  Y &\equiv & \frac{3 \sqrt{\lambda}}{4 \pi \gamma z_H^{3}}\,,\qquad
 M_{Q} \equiv \frac{\sqrt{\lambda} T}{2 z_m}\,=
 \,\frac{\sqrt{\lambda}\,r_m}{2\pi R^2}\,.\eeq
$M_{Q}$ is the (bare) rest mass of the heavy quark and is independent of
temperature, as manifest in his last rewriting. ($r_m$ denotes the
position of the D7--brane in the usual radial coordinate $r$, which is
related to $z$ as $z/\pi T = R^2/r$.) At finite temperature, this mass
receives thermal corrections, as encoded in the contribution of
$\order{\omega^2}$ to $\mbox{Re}\, C_{ret}(\omega)$; such corrections are
however negligible at high energy, since their contribution to
$G_{R}(\omega)$ is not enhanced by a factor of $\gamma$ (unlike $M_Q$).

Note that the previous formulae fully specify the imaginary part of the
retarded propagator, and hence also $G_{\rm sym}(\omega)$. Namely, by
using (cf. Eqs.~(\ref{Cadv})--\eqref{X eq})
 \beq \mbox{Im}\,C_{ret}(\omega)+\mbox{Im}\,C_{adv}^*(\omega)\,=\,
 \frac{2\omega z_H}{3}(1+v^2\gamma^2)\,=\,
 \frac{2\omega z_H}{3}\,\gamma^2,\eeq
one immediately finds
 \beq\label{drag G}
 \mbox{Im}\, G_{R}(\omega)\,=\,-
 \omega\gamma \eta\,, \qquad \mbox{with}\qquad
 \eta \,\equiv\, \frac{\pi \sqrt{\lambda}}{2}\,  T^2
 \,.\eeq
Remarkably, this exact result involves just a term linear in $\omega$. On
the other hand, we expect $\mbox{Re}\, G_{R}(\omega)$ to receive
contributions to all orders in $\omega$ starting at $\order{\omega^2}$.

The above expression for $G_R^0$, \eqnum{GRplicit}, coincides with that
originally derived in Ref.~\cite{Gubser:2006nz}, although the definition
used there for the retarded propagator was different, namely
 \beq\label{GRGubser}
 G^{0}_{R}(\omega)\,\equiv \,- \Psi_{ret}^* T^{z\beta}
\partial_\beta\Psi_{ret} |_{z=z_m}\,.
\eeq
(Recall that $\Psi_{ret} (\omega, t, z) = \rme^{-i \omega
t}\psi_{ret}(\omega, z)$.) With this definition, the dominant
contribution to the imaginary part at high energy --- the term
proportional to $X(\omega)$ --- arises from the time derivative of the
retarded solution. Although it does not naturally emerge when
constructing the boundary action in the Kruskal plane, this formula
\eqref{GRGubser} has another virtue, which will be useful later on: with
this definition, the imaginary part of the retarded propagator,
 \beq\label{ImGR}
 \mbox{Im}\,G_R^0(\omega)\,=\,\frac{1}{2i}\,T^{z\beta}
 \Big(\Psi_{ret}^*\partial_\beta\Psi_{ret} -
 \Psi_{ret}\partial_\beta\Psi_{ret}^*\Big)\,,\eeq
can be evaluated at any $z$, since the r.h.s. of \eqnum{ImGR} is
independent of $z$. Indeed, as noticed in Ref.~\cite{Gubser:2006nz}, the
world--sheet current
 \beq\label{current}
 J^\alpha\,=\,\frac{1}{2i}\,T^{\alpha\beta}
 \Big(\Psi_{sol}^*\partial_\beta\Psi_{sol} -
 \Psi_{sol}\partial_\beta\Psi_{sol}^*\Big)\,,\eeq
($\Psi_{sol}(t,z)$ is an arbitrary solution to the classical EOM
\eqref{EOM}) is conserved by the equations of motion, $\partial_\alpha
J^\alpha=0$. When $\Psi_{sol}(t,z)=\Psi_{ret} (\omega, t, z)$, this
conservation law reduces to $\partial_z J^z=0$. As we shall shortly see,
$\mbox{Im}\,G_R^0(\omega)$ is a measure of the energy loss of the heavy
quark towards the plasma. Thus the fact this quantity is independent of
$z$ is a statement about the conservation of the energy flux down the
string in the present, steady, situation.

\subsection{A Langevin equation for the heavy quark}

Following the general strategy of AdS/CFT, the boundary action \eqref{G
NG} can be used to generate the correlation functions of the
$\mathcal{N}=4$ SYM operator which couples to the boundary value of the
field --- in this case, the Schwinger--Keldysh 2--point functions of the
force operator acting on the heavy quark
\cite{Herzog:2002pc,CasalderreySolana:2007qw,Gubser:2006nz,Son:2009vu}.
Alternatively, in what follows, this action will be used to derive
stochastic equations for the string endpoint, in the spirit of the
Feynman--Vernon `influence functional' \cite{Feynman:1963fq} (see also
Ref.~\cite{Son:2009vu}).


To that aim we start with the following path integral which encodes the
(quantum and thermal) dynamics of the string fluctuations in the Gaussian
approximation of interest:
 \beq\label{Path Int}
  Z = \int \left[ D\psi^{0}_{R} D\psi_{R} \right] \left[ D\psi^{0}_{L}
 D\psi_{L} \right] e^{iS_R - iS_L}\,. \eeq
This involves two types of functional integrations: \texttt{(i)} those
with measure $[D\psi_{R}D\psi_{L}]$, which run over all the string
configurations $\psi_{R,L}(t,z)$ (in the corresponding quadrants of the
Kruskal diagram) with given boundary values $\psi^{0}_{R,L}(t)$, and
\texttt{(ii)} those with measure $[D\psi^{0}_{R}D\psi^{0}_{L}]$, which
run over all the possible paths $\psi^{0}_{R,L}(t)$ for these endpoint
values.

Performing the Gaussian path integral over the bulk configurations
amounts to evaluating the action in the exponent of \eqnum{Path Int} with
the classical solutions computed in the previous section. This leaves us
with the boundary action in \eqnum{G NG}, which determines the dynamics
of the string endpoints --- i.e., of the heavy quark ---, and which is
itself Gaussian. To perform the corresponding path integral it is
convenient to `break' the quadratic term for the fluctuating fields
$\psi_a^0$, by introducing an auxiliary stochastic field $\xi (t)$. Then
the partition function becomes
 \beq\label{Stoch Path Int} Z &=& \int \left[ D\psi^{0}_{r}
 \right] \left[ D\psi^{0}_{a} \right] \left[ D\xi \right]\
 \rme^{-\int \,{\rmd t \rmd t'}\,\frac{1}{2}\, \left[\xi(t)
 G_{\rm sym}^{-1}(t,t')\xi(t') \right]} \nonumber \nn
 &{}& \exp\left\{ -i \int \,{\rmd t \rmd t'}\,
 \psi_a^0 (t) \big[ G_R^0
 (t,t') \psi_r^0(t') +\delta(t-t')\big(P^z - \xi (t')\big)\big] \right\},
 \eeq
where we recall that the term involving $P^z$ appears only in the
longitudinal sector. The integral over $\psi_a$ acts as a constraint
which enforces a Langevin equation for the `average' field $\psi_r$. This
equation reads 
\beq\label{Langevin}
 \int \rmd t'
G_{R}^0(t,t') \psi^{0}_{r}(t')+P^z  - \xi(t)=0,\qquad \langle \xi (t)
\xi(t') \rangle = G_{{\rm sym}}(t,t')\,,
\eeq and is
generally non--local in time. At this point it is convenient to focus on
the large time behaviour, as controlled by the small--frequency expansion
of the Green's functions $G_R$ and $G_{\rm sym}$, and also distinguish
between longitudinal and transverse fluctuations. As discussed in Sect.
2.2, for $\omega\ll z_sT$, the retarded propagator reduces to its
imaginary part, \eqnum{drag G} (in addition to the bare mass term). In
the same limit, one can use $1 + 2 n(\omega)\simeq 2z_sT/\omega$ to
simplify the expression of $G_{\rm sym}(\omega)$, which then becomes
independent of $\omega$\,:
 \beq\label{kappa}
 G_{\perp,\, \rm sym}(\omega)&\,\simeq\,&
 {\pi \sqrt{\lambda}}\,{\gamma}^{1/2}  T^3\,\equiv\,\kappa_\perp
 \,,\nn
 G_{\ell,\, \rm sym}(\omega)&\,\simeq\,&
 {\pi \sqrt{\lambda}}\,\gamma^{5/2} T^3\,\equiv\,
 \kappa_\ell\,.\eeq
This in turn implies that, when probed over large time separations
$t-t'\gg 1/ z_sT$, the retarded propagator can be replaced by a local
time derivative (`friction force'), while the noise--noise correlator
looks local in time (`white noise'). The we can write
 \beq\label{LangT}
\gamma M_{Q} \,\frac{\rmd^{2}\delta x_\perp}{\rmd t^{2}} =  - \gamma \eta
\,\frac{\rmd\delta x_\perp}{\rmd t}+ \xi_\perp(t),\qquad \langle
\xi_\perp(t) \xi_\perp(t') \rangle = \kappa_\perp\delta(t-t')\,, \eeq for
the transverse modes and, respectively (note that $P^z=\gamma\eta v$),
 \beq\label{LangL}
\gamma^3 M_{Q} \,\frac{\rmd^{2}\delta x_\ell}{\rmd t^{2}} =  - \gamma^3
\eta \,\frac{\rmd\delta x_\ell}{\rmd t}\,- \gamma\eta v +
\xi_\ell(t),\qquad \langle \xi_\ell (t)
 \xi_\ell(t') \rangle = \kappa_\ell\,\delta(t-t')\,, \eeq
for the longitudinal one. The physical interpretation of these equations
becomes more transparent if they are first rewritten in terms of the
respective momenta $p_\perp=\gamma M_Qv_\perp$ and $p_\ell=\gamma
M_Qv_\ell$, with $v_\perp={\rmd\delta x_\perp}/{\rmd t}$ and $v_\ell=
v+{\rmd\delta x_\ell}/{\rmd t}$.

At this point, we come across a rather subtle point: in all the equations
written so far, the Lorentz factor $\gamma$ is evaluated with the {\em
average} velocity $v$ of the heavy quark --- the one which enters the
trailing string solution \eqref{Trail}. However, the event--by--event
fluctuations of the velocity turn out to be significantly large
(especially in the longitudinal sector; see below), and then it becomes
appropriate to define the {\em event--by--event} (or `fluctuating')
momenta $p_\perp$ and $p_\ell$ by using the respective, event--by--event,
Lorentz factor, as evaluated with the instantaneous velocity. For more
clarity, let us temporarily denote by $v_0$ and $\gamma_0$ the average
velocity and the associated Lorentz factor, $\gamma_0\equiv
1/\sqrt{1-v_0^2}$, and reserve the notations $v$ and $\gamma$ for the
respective fluctuating quantities:
 \beq
 v^2 = v^2_\ell+v^2_\perp=\left(v_0 + \frac{\rmd\delta x_\ell}{\rmd t}
 \right)^2
 + \left(\frac{\rmd\delta x_\perp}{\rmd t}\right)^2\,,\qquad
 \gamma\,=\,\frac{1}{\sqrt{1-v^2}}\,.
 \eeq
When taking the time derivatives of $p_\perp$ and $p_\ell$, as associated
with variations in $v_\perp$ and, respectively, $v_\ell$, one must also
take into account the corresponding change in the $\gamma$--factor.
Consider the longitudinal sector first:
 \beq\label{dpl}
 \frac{1}{M_Q}\frac{\rmd p_\ell}{\rmd t}=\left(\gamma
 +v_\ell\frac{\del\gamma}{\del v_\ell}\right)\frac{\rmd v_\ell}{\rmd t}
 =\left(\gamma + v_\ell^2\gamma^3  \right)
 \frac{\rmd^2\delta x_\ell}{\rmd t^2}
 \simeq \gamma_0^3\,\frac{\rmd^2\delta x_\ell}{\rmd t^2}\,,\eeq
where the last, approximate, equality follows since the fluctuations are
assumed to be small, hence $v_\ell\simeq v_0$ and $\gamma\simeq\gamma_0$.
The final result above is recognized as the expression in the l.h.s. of
\eqnum{LangL}. To the same accuracy, we can write (with $\delta v_\ell
={\rmd\delta x_\ell}/{\rmd t}$)
 \beq \gamma v_\ell\simeq \left(\gamma_0+
 \frac{\del\gamma}{\del v_\ell}\,\delta v_\ell\right)(v_0 + \delta v_\ell)
 \simeq \gamma_0v_0+(\gamma_0+\gamma_0^3v_0^2)\delta v_\ell
 = \gamma_0v_0 + \gamma_0^3 \delta v_\ell\,,
 \eeq
in which we recognize the terms multiplying $\eta$ in the r.h.s. of
\eqnum{LangL}.

Consider similarly the transverse sector. The analog of \eqnum{dpl} reads
 \beq\label{dpt}
 \frac{1}{M_Q}\frac{\rmd p_\perp}{\rmd t}=\left(\gamma
 +v_\perp\frac{\del\gamma}{\del v_\perp}\right)\frac{\rmd v_\perp}{\rmd t}
 =\gamma\left(1 + v_\perp^2\gamma^2  \right)
 \frac{\rmd^2\delta x_\perp}{\rmd t^2}\simeq
 \gamma_0 \frac{\rmd^2\delta x_\perp}{\rmd t^2}\,,\eeq
where we made the additional assumption that $v_\perp^2\ll 1-v_0^2$.
(This can be always ensured by taking the quark mass $M_Q$ to be
sufficiently large.) Similarly, in the r.h.s. of \eqnum{LangT}, one can
replace $\gamma_0 v_\perp \simeq p_\perp/M_Q$.

To summarize, to the accuracy of interest, we have derived the following
Langevin equations for the dynamics of the heavy quark
 \beq\label{LangPT}
 \frac{\rmd p_\perp^i}{\rmd t}&=&-\eta_D p_\perp^i
 + \xi_\perp^i(t),\qquad \langle
\xi_\perp^i(t) \xi_\perp^j(t') \rangle = \kappa_\perp\delta^{ij}
 \delta(t-t')\,,\\
 \frac{\rmd p_\ell}{\rmd t}&=&-\eta_D p_\ell+
  \xi_\ell(t),\qquad \ \langle
 \xi_\ell (t)
 \xi_\ell(t') \rangle = \kappa_\ell\,\delta(t-t')\,,
 \label{LangPL}\eeq
where the upper index $i=1,\,2$ in \eqnum{LangPT} distinguishes between
the two possible transverse directions, $\kappa_\perp$ and $\kappa_\ell$
are given in \eqnum{kappa}, and
 \beq\label{etaD}
 \eta_D \,\equiv\,  \frac{\eta}{M_Q}\,=\,
 \frac{\pi \sqrt{\lambda}}{2M_Q}\, T^2\,.\eeq
The general structure of these equations --- with a friction term (or
`drag force') describing dissipation and a noise term describing momentum
broadening --- is as expected, and so are the above expressions for
$\eta_D$, $\kappa_\perp$ and $\kappa_\ell$, which agree with previous
calculations in the literature
\cite{CasalderreySolana:2006rq,CasalderreySolana:2007qw,Gubser:2006nz}.
It is however important to keep in mind that
Eqs.~(\ref{LangPT})--(\ref{etaD}) have been derived here only for the
situation where the fluctuations in the velocity of the heavy quark
remain small as compared to its average velocity $v_0$. To ensure that
this is indeed the case, \eqnum{LangPL} for the longitudinal motion must
be supplemented with an external force which is tuned to reproduce the
average motion. (Without such a term, \eqnum{LangPL} would describe the
rapid deceleration of the heavy quark due to its interactions in the
plasma. Such a deceleration may entail additional phenomena, like
bremsstrahlung, which are not encoded in the above equations; see the
discussion in Refs.~\cite{Dominguez:2008vd,Xiao:2008nr,Fadafan:2008bq}.)
Namely, we shall add to the r.h.s. of \eqnum{LangPL} a term
$F_{ext}=\eta\gamma_0v_0$ which for large times equilibrates the average
friction force and thus enforces a constant average velocity $v_0$.
Further consequences of these equations will be discussed in Sect. 4.

\section{Bulk picture of the stochastic motion}
\setcounter{equation}{0}

In the previous section we have obtained a set of Langevin equations for
the heavy quark by integrating out the fluctuations of the upper part of
the string, from the world--sheet horizon up to the boundary. The noise
terms in these equations have been generated via boundary conditions at
the world--sheet horizon, cf. \eqnum{Modes R to L}. This suggests that,
within the context of the supergravity calculation, quantum fluctuations
are somehow encoded in the world--sheet horizon. To make this more
explicit, we shall follow Refs.~\cite{Son:2009vu,deBoer:2008gu} and
construct a set of equations describing the stochastic dynamics of the
upper part of the string, in which the noise term is acting on the {\em
lower} endpoint, infinitesimally close to the world--sheet horizon.

More precisely, we introduce a `stretched' horizon at $z_h\equiv
z_s-\epsilon$ and integrate out the fluctuations of the part of string
lying between $z_s$ and $z_h$. The procedure is quite similar to the one
described in the previous section except that one has to fix the boundary
values for the fluctuations also on the stretched horizon, rather than
just on the Minkowski boundary. Denoting the respective values by
$\psi^h$, where as before $\psi$ stands generically for either $\delta
x_\ell$ or $\delta x_\perp$, this procedure yields an effective action
$S^{h}_{\rm eff}$ for $\psi^h$ with the same formal structure as
exhibited in Eq.~(\ref{G NG}), that is,
 \beq\label{Wrldsht actn}
i S^{h}_{\rm eff} = - i \int \frac{\rmd\omega}{2 \pi} \psi^{h}_a (-
\omega) G^{h}_R (\omega)  \psi^{h}_r (\omega) - \frac{1}{2} \int
\frac{\rmd\omega}{2 \pi} \psi^{h}_a (- \omega) G^{h}_{\rm sym}(\omega)
 \psi^{h}_a (\omega). \eeq
(We temporarily omit the term linear in the fluctuations; this will be
restored in the final equations.) The horizon Green's functions $G^{h}_R$
and $G^{h}_{\rm sym}$ will be shortly constructed. The calculations being
quite involved, it is convenient to start with a brief summary of our
main results:

The $r$--fields $\psi_r(\omega, z)$ describing the string fluctuations
within the bulk ($z_m\le z\le z_h$) obey the equations of motion
\eqref{EOM} with Neumann boundary conditions at $z=z_m$ --- meaning that
the string endpoint on the boundary is freely moving (except for the
imposed longitudinal motion with velocity $v_0$) --- and with Dirichlet
boundary conditions at $z=z_h$: $\psi_r(\omega, z_h)=\psi_r^h(\omega)$.
This boundary field $\psi_r^h(\omega)$ is however a stochastic variable,
whose dynamics is described by the effective action \eqref{Wrldsht actn}.
Via the classical solutions, this stochasticity is transmitted  to the
upper endpoint of the string, i.e., to the heavy quark. As a result, the
latter obeys the same Langevin equations as previously derived in Sect.
2.

We start with the partition function encoding the quantum dynamics of the
upper part of the string ($z_m\le z\le z_h$) in the Gaussian
approximation:
\beq\label{Hrzn Pth}
Z = \int \left[ D\psi^{0}_{R} D\psi_{R} D\psi_{R}^{h} \right] \left[
D\psi^{0}_{L} D\psi_{L} D\psi_{L}^{h} \right] \,\rme^{iS_R - iS_L + i
 S^{h}_{\rm eff} }. \eeq
The different measures $D\psi^{0}$, $D\psi$ and $D\psi^{h}$ correspond,
respectively, to the path integral over the string endpoint on the
Minkowski boundary, over the bulk part of the string, and over the string
endpoint on the stretched horizon (separately for the left and right
quadrants of the Kruskal plane). Also, $S_R$ and $S_L$ are defined as in
\eqnum{NG}, but with the integral over $z$ restricted to $z_m<z<z_h$. In
what follows we shall construct the various pieces of the action which
enter the exponent in \eqnum{Hrzn Pth}.

{\sf (I) The effective action at the stretched horizon, $S^{h}_{\rm
eff}$.} As anticipated, this is obtained by integrating out the string
fluctuations within the infinitesimal strip $z_h<z<z_s$. To that aim, we
need the classical solutions $\psi_{R}(\omega, z)$ and $\psi_{L}(\omega,
z)$ in the Kruskal plane which take the boundary values
$\psi_{R}^{h}(\omega)$ and $\psi_{L}^{h}(\omega)$ at $z=z_h$ and are
related by the condition \eqref{Modes R to L}. Clearly, the respective
solutions read (in the $(r,a)$ basis, for convenience)
 \beq\label{psih}
\psi_{r}(\omega, z) & =& \psi^{h}_{r}(\omega) \psi_{ret}^h(\omega, z) +
\frac{1+2n(\omega)}{2}\, \psi^{h}_{a}(\omega) (\psi_{ret}^h(\omega, z) -
\psi_{adv}^h(\omega, z)),\nn
 \psi_{a}(\omega, z) &=& \psi^{h}_{a}(\omega)
 \psi_{adv}^h(\omega, z),
 \eeq
where $\psi_{ret}^h$ and $\psi_{adv}^h$ are rescaled versions of the
retarded and advanced solutions introduced in Sect. 2.2 which are
normalized to 1 at $z=z_h$; e.g., $\psi_{ret}^h(\omega, z)=
\psi_{ret}(\omega, z)/\psi_{ret}(\omega, z_h)$. For $z$ close to $z_s$
(and hence to $z_h$ as well), these functions can be expanded as
 \beq\label{psi -}
 \psi_{ret}^h(\omega, z) &=& 1 + O(z_s - z)\,,\nn
 \psi_{adv}^h(\omega, z)&=&\left(\frac{z_s-z}{z_s-z_h}
 \right)^{\frac{i \omega z_H}{2
 z_s}}\left[1+\order{z_s-z}\right]\,.\eeq
Substituting these classical solutions into the action produces the
boundary action shown in \eqnum{Wrldsht actn}, with $G^{h}_R$ defined by
the horizon version of \eqnum{G_R}. Given the near--horizon behaviour of
the solutions \eqref{psi -} and of the local tension $T^{zz}(z)$ (which
vanishes at $z=z_s$, cf. \eqnum{T pert}), it is clear that only
$\partial_z\psi_{adv}^h$ contributes to $G^{h}_R$ in the limit
$\epsilon\to 0$. This yields the following, purely imaginary, result:
\beq\label{G R h}
 G^{h}_{\perp,R} (\omega)\,=\, z_s^{4} G^{h}_{\ell,\, R}(\omega)
 \,=\,- \frac{1}{2}
 T^{zz}_{\perp}(z_h)\partial_z \psi_{adv}^h(-\omega, z)\big|_{z= z_h}
 \,= \,
 -i\omega\gamma \eta\,. \eeq
This coincides, as it should, with the imaginary part of the respective
boundary propagator\footnote{Incidentally, this calculation of ${\rm
Im}\,G^{h}_R$, which is exact, together with the conservation law
$\partial_z J^z=0$, cf. \eqnum{current}, can be used to check, or even
derive, the expressions for $\mbox{Im}\,C_{ret}(\omega)$ and
$\mbox{Im}\,C_{adv}(\omega)$ given in Eqs.~(\ref{Cadv})--\eqref{X eq}.},
\eqnum{drag G} (cf. the discussion at the end of Sect. 2.2). Then the
symmetric Green's function $G^{h}_{\rm sym}$, which is related to ${\rm
Im}\,G^{h}_R$ via the KMS relation (\ref{G_sym}), is exactly the same as
the corresponding function on the boundary.

If the string point on the stretched horizon was a free endpoint endowed
with the action \eqref{Wrldsht actn}, it would obey Langevin equations
similar to those derived in Sect. 2.3. However, this is an internal point
on the string, and as such it is also subjected to a tension force from
the upper side of the string at $z<z_h$. This force is encoded in the
bulk action $S_R - S_L$, to which we now turn.

{\sf (II) The bulk piece of the action $S_R - S_L$.} This is defined as
 \beq\label{SRLout}
S_R - S_L&=&- \frac{1}{2} \int_{z_m}^{z_h}\rmd z\int\rmd t\,
 T^{\alpha
\beta}(z)\Big[\partial_{\alpha}\psi_R\partial_{\beta}\psi_R
- \partial_{\alpha}\psi_L\partial_{\beta}\psi_L\Big]\nn
  &=&\frac{1}{2} \int_{z_m}^{z_h}\rmd z\int\rmd t\,
\Big[\psi_R\partial_{\alpha} \Big(T^{\alpha
\beta}\partial_{\beta}\psi_R\Big) -\psi_L
\partial_{\alpha}\Big(T^{\alpha
\beta}\partial_{\beta}\psi_L\Big)\Big]\nn
 &{}&- \frac{1}{2}\int\rmd t\,T^{zz}(z)\Big(\psi_R\partial_z \psi_R
 -\psi_L\partial_z \psi_L\Big)\bigg|_{z= z_m}^{z= z_h}
 \eeq
or, after going to Fourier space and to the $(r,a)$--basis,
 \beq\label{SRLoutF}
S_R - S_L&=& \int \frac{\rmd\omega}{2 \pi} \int \rmd z\, \psi_a (-\omega,
z)
\partial_{\alpha}\left[ T^{\alpha
\beta} (z)
\partial_{\beta} \psi_r(\omega, z)\right]\nn
 &{}&-\frac{1}{2} \int \frac{\rmd\omega}{2 \pi} \,{T^{zz}(z)}
 \Big( \psi_a (-\omega, z) \partial_z \psi_r(\omega, z) +
 (r\leftrightarrow a) \Big)\bigg|_{z= z_m}^{z= z_h}.
 \eeq
We were so explicit here about the integration by parts, because this
operation turns out to be quite subtle. First, notice that the
contributions proportional to $T^{zt}$ have cancelled in the boundary
terms, for the same reason as discussed below \eqnum{bndry NG}, i.e.,
because they are odd functions of $t$ (or $\omega$). To ensure this
property, it has been important to perform the previous operations in the
order indicated above, that is, to first integrate by parts, as in
\eqnum{SRLout}, and only then change to the $(r,a)$--basis, as in
\eqnum{SRLoutF}. (Reversing this order would have affected the symmetry
properties of the integrand, and then the terms $\propto T^{zt}$ would
not cancel anymore.)

Second, there are some subtleties about the boundary value of $S_R-S_L$
at the stretched horizon, that we rewrite here for more clarity:
 \beq\label{SRLh}
(S_R - S_L)^h_{\rm bndry}&=&
 -\frac{1}{2} \int \frac{\rmd\omega}{2 \pi} \,{T^{zz}(z_h)}
 \Big( \psi_a (-\omega, z) \partial_z \psi_r(\omega, z) +
 (r\leftrightarrow a) \Big)\bigg|_{z= z_h}
 \eeq
If we were to evaluate this action with the classical solutions
\eqref{psih}, the result would precisely cancel the effective action
\eqref{Wrldsht actn} in the exponent of \eqnum{Hrzn Pth}. Indeed, up to a
sign, \eqnum{SRLh} has exactly the structure that has been used to build
the effective action by inserting the classical solutions (compare to
\eqnum{G NG0}). However, in the present context, \eqnum{SRLh} must be
rather seen as the boundary value of the bulk action when approaching the
stretched horizon from the {\em above} (i.e., from $z<z_h$), and as such
it provides boundary conditions for the dynamics of the upper side of the
string (see below). This being said, it is nevertheless possible, and
also convenient, to use the appropriate piece of \eqnum{SRLh} in order to
cancel the dissipative piece $\propto G^{h}_R$ in the effective action
\eqref{Wrldsht actn}. This is simply the statement that the totality of
the energy which crosses the stretched horizon coming from the above
flows further down along the string.

Specifically, the relevant piece of \eqnum{SRLh} is that proportional to
$\partial_z \psi_{a}$, which after using \eqnum{psih} can be evaluated as
 \beq\label{deradv}
 -\frac{1}{2}
 T^{zz}(z_h)\partial_z \psi_{adv}^h(-\omega, z)\big|_{z= z_h}
 \,=\,G^{h}_{R} (\omega)
 \eeq
where we have recognized the expression \eqref{G R h} for $G^{h}_R$. One
thus obtains:
  \beq\label{SRLh3}
(S_R - S_L)^h_{\rm bndry}&=&
 - \int \frac{\rmd\omega}{2 \pi} \,\,
 \psi_a^{h}(-\omega)\left[\frac{1}{2}\,
 T^{zz}(z)\partial_z \psi_r(\omega,z)
 - G^{h}_{R} (\omega) \psi_r^{h}(\omega)
 \right]_{z= z_h}.\eeq
As anticipated, the last term in \eqnum{SRLh3} compensates the piece
involving $G^{h}_R$ in \eqnum{Wrldsht actn}, and then the total action
reads
 \beq\label{SRLH}
iS_R - iS_L + i S^{h}_{\rm eff}& =& i  \int \frac{\rmd \omega}{2 \pi}
\,\psi_a^{0}(-\omega) \left[\frac{1}{2}\, T^{zz}(z)\Big(
\partial_z \psi_r(\omega,z)+ \psi_r^{0}(\omega)
 \partial_z \psi_{adv}(-\omega, z)\Big) - P^{z} \right]_{z= z_m}
\nonumber \\
 & + &i  \int_{z_m}^{z_h}
 \rmd z \int \frac{\rmd\omega}{2 \pi}\, \psi_a (-\omega, z)
\partial_{\alpha}\left[ T^{\alpha
\beta} (z)
\partial_{\beta} \psi_r(\omega, z)\right] \nonumber \\
& -& i \int \frac{\rmd \omega}{2 \pi}\, \psi_a^{h}(-\omega)\left[
\frac{1}{2}\,T^{zz}(z)
\partial_z \psi_r(\omega,z) - \xi^{h}(\omega) \right]_{z= z_h} \eeq
where the differences between longitudinal and transverse fluctuations
are kept implicit (in particular, it is understood that the term
proportional to $P^z$ appears only in the longitudinal sector). Two
additional manipulations have been necessary to write the action in its
above form: \texttt{(i)} In the first line of \eqnum{SRLH} we have used
$\psi_{a}(\omega, z) = \psi^{0}_{a}(\omega) \psi_{adv}(\omega, z)$, cf.
\eqnum{psi r}. \texttt{(ii)} The piece involving $G^{h}_{\rm
sym}(\omega)$ in \eqnum{Wrldsht actn} have been reexpressed as a Gaussian
path integral over the noise variables $\xi^h$, which therefore obey
 \beq\label{noiseh}\big\langle \xi^{h}(\omega)
 \xi^{h}(\omega') \big\rangle \,=\, 2\pi\delta(\omega+\omega')
 \,(1 + 2 n) \omega\gamma\eta
 \,,\quad\mbox{with}
 \quad n(\omega) = \frac{1}{\rme^{\omega/z_s T}-1}\,.\eeq
Once again, \eqnum{noiseh} involves the effective temperature $T_{\rm
eff} = z_s T$.

By integrating over the fluctuating fields $\psi_a^{0}(-\omega)$, $\psi_a
(-\omega, z)$, and $\psi_a^{h}(-\omega)$, we are finally left with the
following set of equations of motion and boundary conditions:

\texttt{(1)} A modified Neumann boundary condition for the string
endpoint string at the boundary (we temporarily reintroduce the
polarization label $p$ with $p=\ell$ or $\perp$) :
\beq\label{Neumann}
 \frac{1}{2}\,T^{zz}_{p}(z)\Big[
\partial_z \psi_r^p(\omega,z)+ \psi_r^{0,p}(\omega)
 \partial_z \psi_{adv}(-\omega, z)\Big]_{z= z_m}
 \,=\, \eta\gamma v\, \delta_{p\ell}\,. \eeq

\texttt{(2)} The standard equations of motion for the fluctuations
$\psi_r(\omega, z)$ of the string in the bulk at $z_m < z <z_h$ (cf.
\eqnum{EOM}).

\texttt{(3)}  A stochastic equation for the string endpoint on the
stretched  horizon :
\beq\label{hend}
 \frac{1}{2}\,
T^{zz}(z) \partial_z \psi_r(\omega,z)\big |_{z= z_h} \,= \,
\xi^{h}(\omega).
\eeq

We now analyze the consequences of these equations and, in particular,
emphasize the differences w.r.t. the corresponding analysis in
Ref.~\cite{Son:2009vu}.

\eqnum{hend} is a Langevin equation of a special type: the noise term is
precisely compensating the pulling force $T^{zz}(z_h) \partial_z \psi_r$
due to the string tension. By taking the expectation value of this
equation and recalling that $T_{p}^{zz}(z_h)\sim\epsilon$ vanishes in the
limit $\epsilon\to 0$, we conclude that $\partial_z
\langle\psi_r^{p}(\omega,z)\rangle$ is regular near the world--sheet
horizon; this implies that the average value of the classical solution is
proportional to $\psi_{ret}$ :
 \beq\label{psirav}
 \langle\psi_{r}(\omega, z)\rangle \, =\,
  \langle\psi^{0}_{r}(\omega) \rangle\,\psi_{ret}(\omega, z)\,.\eeq
The normalization is fixed by the expectation value of
$\psi^{0}_{r}(\omega)$ --- the boundary value of $\psi_{r}(\omega, z)$ at
$z=z_m\ll 1$.

We will now construct the solution $\psi_r(\omega, z)$ to the EOM
\eqref{EOM} by specifying its boundary values, $\psi^{0}_{r}(\omega)$ and
$\psi^{h}_{r}(\omega)$, at the endpoints $z=z_m$ and $z=z_h$,
respectively. After simple algebra, the respective solution can be
written as
 \beq\label{psircl}
\psi_{r}(\omega, z) & =& \psi^{0}_{r}(\omega) \psi_{ret}(\omega, z) +
\big[\psi_r^h(\omega)-\psi^{0}_{r}(\omega)\psi_{ret}(\omega,z_h)\big] \,
 \frac{\psi_{ret}(\omega, z) - \psi_{adv}(\omega, z)}
 {\psi_{ret}(\omega,z_h)-\psi_{adv}(\omega,z_h)}\,.\nn
 \eeq
The reason why this particular writing is natural is as follows: when
taking the expectation value according to \eqnum{psirav}, we find
$\langle\psi_{r}^h(\omega)\rangle  = \langle\psi^{0}_{r}(\omega)
\rangle\,\psi_{ret}(\omega, z_h)$, which shows that the coefficient
$\psi_r^h(\omega)-\psi^{0}_{r}(\omega)\psi_{ret}(\omega,z_h)$ in front of
the second term in \eqnum{psircl} is a random variable with zero
expectation value. Clearly, this term plays the role of a noise. The
statistics of this noise is determined by the horizon Langevin equation
\eqref{hend}, and in turn it implies a boundary Langevin equation for
$\psi^{0}_{r}(\omega)$, via the condition \eqref{Neumann}. Let's see how
all that works in detail. We will first rewrite \eqnum{psircl} as
 \beq\label{psirxi}
\psi_{r}(\omega, z) & =& \psi^{0}_{r}(\omega) \psi_{ret}(\omega, z) + i\,
\xi^0(\omega)\frac{\psi_{ret}(\omega, z) - \psi_{adv}(\omega, z)}
 {{\rm Im} \,G_R(\omega)}\,,
 \eeq
thus fixing the normalization of the noise term $\xi^0(\omega)$. After
inserting \eqnum{psirxi} in the Neumann boundary condition
\eqref{Neumann} (say, in the transverse sector), one finds\footnote{The
following identities, which can be checked from \eqnum{G_R}, are useful
in this respect:
$$G^{0}_{\perp,\, R}(\omega)= - \frac{1}{2}\, T_{\perp}^{zz}(z_m)\,\big[
\partial_z\psi_{ret}(\omega,z)
 +\partial_z\psi_{adv}(-\omega,z)\big]_{z= z_m}.$$
$${\rm Im} \,G_R(\omega)=\frac{i}{2}\,
T_{\perp}^{zz}(z_m)\,\partial_z\big[\psi_{ret}(\omega,z)
-\psi_{adv}(\omega,z)\big]_{z= z_m}.$$}
 \beq\label{Langevin2}
 G_R^0(\omega)\,\psi^{0}_{r}(\omega)\,=\,\xi^0(\omega)\,,\eeq
which is the standard form of a Langevin equation (compare to
\eqnum{Langevin}).

It remains to check that the statistics of $\xi^0$, as inferred from
\eqnum{hend}, is indeed the same as previously derived in Sect. 2.3. To
that aim, we insert the form \eqref{psirxi} of the solution into
\eqnum{hend}; as already explained, the regular piece of the solution
$\propto \psi_{ret}(\omega, z)$ does not contribute in the limit
$\epsilon\to 0$, so we are left with
 \beq\label{Langhor}
 -\frac{i}{2}\,\frac{\xi^0(\omega)}{{\rm Im} \,G_R(\omega)}\,
T^{zz}(z) \partial_z\psi_{adv}(\omega, z)\big|_{z= z_h}
 \,= \,\xi^{h}(\omega).
 \eeq
Using $T^{zz}(z) \partial_z\psi_{adv}(\omega, z)|_{z= z_h} =-2i
\omega\gamma \eta \psi_{adv}(\omega, z_h)$, cf. \eqnum{deradv}, together
with ${\rm Im} \,G_{R}(\omega)= -  \omega\gamma \eta$, this finally
becomes
 \beq\label{xi0h}
 \psi_{adv}(\omega, z_h)\,\xi^0(\omega) \,= \,\xi^{h}(\omega)\,.
 \eeq
This relation is in fact natural, as we argue now: from the
transformation connecting $\xi$ to $\psi_a$, or directly by comparing
\eqnum{psirxi} with the standard expression \eqref{psi r} for $\psi_{r}$,
one can see that the strength of the noise term scales like
$\xi(\omega)\sim \psi_a(\omega)\,G_{\rm sym}(\omega)$. On the other hand,
\eqnum{psi r} implies $\psi^{h}_{a}(\omega)= \psi_{adv}(\omega,
z_h)\psi^{0}_{a}(\omega)$. Hence one can write
 \beq
 \xi^{h}(\omega)\,\sim\, \psi_a^h(\omega)\,G_{\rm sym}^h(\omega)
 \,\simeq\, \psi_{adv}(\omega, z_h)\psi^{0}_{a}(\omega)
 \,G_{\rm sym}(\omega)\,\sim\, \psi_{adv}(\omega, z_h)
 \,\xi^0(\omega)\,.\eeq
Remarkably, the relative factor between $\xi^0$ and $\xi^{h}$ in
\eqnum{xi0h} does not spoil the normalization of the noise--noise
correlator, because $|\psi_{adv}(\omega, z_h)|=1$, as we now demonstrate.
To that aim we rely on the observation at the end of Sect. 2.2 that the
r.h.s. of \eqnum{ImGR} is independent of $z$. Clearly, this remains true
after replacing $\psi_{ret}\to \psi_{adv}$ in \eqnum{ImGR}. (Indeed, the
current \eqref{current} is conserved for an arbitrary solution
$\Psi_{sol}(t,z)$.) Writing $\psi_{adv}(\omega,
z)=C(\omega)\psi_{adv}^h(\omega, z)$, so that $\psi_{adv}(\omega,
z_h)=C(\omega)$, and evaluating the r.h.s. of \eqnum{ImGR} separately at
$z=z_m$ and $z=z_h$, one deduces that $|C(\omega)|=1$, as anticipated.
Thus, the 2--point function $\langle \xi^0(\omega)
\xi^{0}(\omega')\rangle$ is indeed the same as in Sect. 2.3. Note also
that our previous argument is independent of the precise value of
$\epsilon$ (the distance between the world--sheet and the stretched
horizons), so long as $\epsilon$ is small enough for the near--horizon
expansions to make sense. This strongly suggests that the strength of the
noise remains constant along the string, from the stretched horizon up to
the boundary.

\section{Discussion and physical picture}
\setcounter{equation}{0}

In this section, we will first discuss some consequences of the
previously derived Langevin equations, which support the idea that the
noise terms in these equations are of non--thermal nature, and then
propose a physical picture in which these fluctuations are interpreted as
quantum mechanical fluctuations associated with medium--induced
radiation.

\subsection{Momentum distributions from the Langevin equations}

An important property of the Langevin equations
(\ref{LangPT})--(\ref{LangPL}) that we would like to emphasize is that,
except in the non--relativistic limit $\gamma\simeq 1$, these equations
do not describe the thermalization of the heavy quark. There are several
arguments to support this conclusion. For instance, in thermal
equilibrium the momentum distributions should be isotropic, but this is
clearly not the case for the large--time distributions generated by
Eqs.~(\ref{LangPT})--(\ref{LangPL}), because of the mismatch between
$\kappa_\ell$ and $\kappa_\perp$ when $\gamma > 1$. Besides, in order to
generate the canonical distribution for a relativistic particle,
$P\propto \exp\{-\sqrt{{\bf p}^2+M_Q^2}/T\}$, the noise correlations must
not only be isotropic, but also obey the relativistic version of the
Einstein relation, which reads $\kappa=2ET\eta_D$ \cite{Mallick97}. Using
Eqs.~\eqref{kappa} and \eqref{etaD}, it is easily seen that this
condition is not satisfied for either transverse, or longitudinal,
fluctuations (except if $\gamma=1$, once again).

The Einstein relation is a particular form of the
fluctuation--dissipation theorem, so its failure might look surprising
given that the Green's functions at the basis of our Langevin equations
obey the KMS condition \eqref{G_sym}. Recall, however, that this peculiar
KMS condition involves an effective temperature $T_{\rm eff} =
T/{\gamma}^{1/2}$; and indeed, in the transverse sector at least, the
Einstein relation appears to be formally satisfied, but with $T\to T_{\rm
eff}$. But this does not hold in the longitudinal sector, where
$\kappa_\ell$ involves an additional factor $\gamma^2$. Hence, the
present equations cannot lead to thermal distributions.

It is then interesting to compute the actual momentum distributions
generated by these Langevin equations at large times. Consider first the
transverse sector, and introduce the probability distribution
$P(\bm{p}_\perp,t)$ for the transverse momentum $\bm{p}_\perp=(p_1,p_2)$
at time $t$:
 \beq\label{Pdef}
 P(\bm{p}_\perp,t)\,\equiv\,\int [D\xi_i]\,\delta\Big(\bm{p}_\perp-
 \bm{p}_\perp[\xi_i](t)\Big)\,\rme^{-\frac{1}{2\kappa_\perp}
 {\int\rmd t \,\xi_i(t)\xi_i(t)}}\,.
 \eeq
Here, $\bm{p}_\perp[\xi_i](t)$ is the solution to \eqnum{LangPT}
corresponding to a given realization of the noise\footnote{In general,
this solution will depend on our prescription for discretizing the time
axis; this is so since the noise--noise correlator depends itself on the
momentum (`multiplicative noise'). But to the accuracy of interest, we
can treat $\gamma$ in \eqnum{kappa} as the fixed quantity $\gamma_0$, and
then one can safely use continuous notations.}, and reads (with
$i=1,\,2$; we assume $p_i(0)=0$ so that $\langle
\bm{p}_\perp(t)\rangle=0$ at any time)
 \beq\label{pT}
 p_i(t)\,=\,\int_0^t\rmd t'\,\rme^{-\eta_D(t-t')}\,\xi_i(t')\,.
 \eeq
This implies $\langle p_1^2\rangle=\langle p_2^2\rangle
 \equiv \langle {p}_\perp^2\rangle$ with
 \beq
 \langle {p}_\perp^2(t)\rangle\,=\,\frac{\kappa_\perp}{2\eta_D}
 \left(1-\rme^{-2\eta_Dt}\right)\,\simeq\,\frac{\kappa_\perp}{2\eta_D}
 \,,\eeq
where the last, approximate equality holds for large times $\eta_D t\gg
1$. Returning to \eqnum{Pdef}, this gives
 \beq
 P(\bm{p}_\perp,t)\,=\,\frac{1}
 {2\pi\langle p_\perp^2(t)\rangle}
 \,\exp\left\{-\frac{p_1^2+p_2^2}{2\langle p_\perp^2(t)\rangle}\right\}.
 \eeq
A similar expression holds in the longitudinal sector, but only after
subtracting away the global motion with velocity $v_0$, which one can do
by writing $\delta p_\ell\equiv p_\ell - p_0$ with $p_0=M_Q\gamma_0 v_0$.

For large times $t\gg 1/\eta_D$, the transverse and longitudinal momentum
distributions for the heavy quark approach the following, stationary,
forms
  \beq
 P(\bm{p}_\perp,t)&\,\simeq\,&\frac{1}{{2\pi {\gamma}^{1/2} TM_Q}}
 \,\exp\left\{-\frac{p_1^2+p_2^2}{2 {\gamma}^{1/2} TM_Q}\right\}\,,\nn
 P(\delta p_\ell\,,t)&\,\simeq\,&\frac{1}{\sqrt{2\pi {\gamma}^{5/2} TM_Q}}
 \exp\left\{-\frac{\delta p_\ell^2}{2 {\gamma}^{5/2} TM_Q}\right\}\,,
 \eeq
which formally look like thermal, Maxwell--Boltzmann, distributions for
{\em non--relativistic} particles, but with different temperatures in the
transverse and longitudinal sector --- $T_\perp = {\gamma}^{1/2} T$ and
$T_\ell={\gamma}^{5/2} T$ ---, none of them equal to the plasma
temperature $T$.

\subsection{Physical picture: Medium--induced radiation}

In this section, we propose a physical picture for the dynamics of the
heavy quark, as encoded in the Langevin equations
(\ref{LangPT})--(\ref{etaD}). The general picture is that of
medium--induced parton branching, as previously developed in
Refs.~\cite{Hatta:2007cs,Hatta:2008tx,Dominguez:2008vd,Iancu:2008sp},
that we shall here adapt to the problem at hand. As we will see, this
qualitative and admittedly crude picture provides the right parametric
estimates for both the drag force and the (transverse and longitudinal)
momentum broadening. Besides, it supports the non--thermal nature of the
noise terms in the Langevin equations.

\begin{figure}
\centerline{
\includegraphics[width=0.7\textwidth]{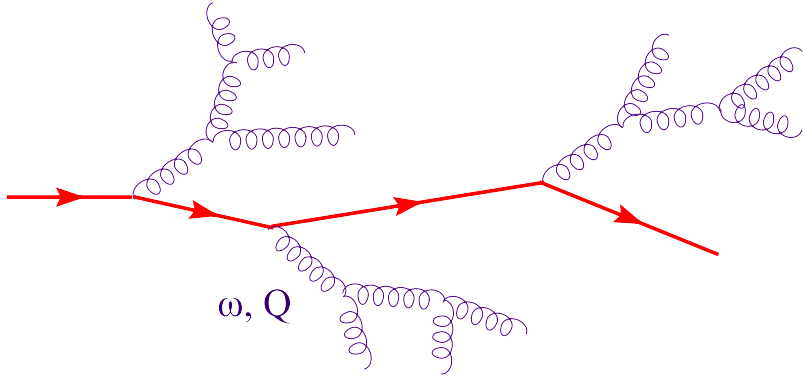}
}
\caption{\sl Energy loss and momentum broadening via medium--induced
parton emission at strong coupling. It is understood that the radiated
partons feel a plasma force which allows them to be
liberated from the parent heavy quark (see text for details).
\label{fig:Broad}}\end{figure}

Due to its interactions with the strongly--coupled plasma, a heavy quark
can radiate massless ${\mathcal N}=4$ SYM quanta (gluons, adjoint scalars
and fermions) which then escape in the medium, thus entailing energy loss
towards the plasma and momentum broadening (due to the recoil of the
heavy quark associated with successive parton emissions). This dynamics
is illustrated in Fig.~\ref{fig:Broad}. The main ingredients underlying
our physical picture are as follows:

\texttt{(i)} The emission of a virtual parton with energy $\omega$ and
(space--like) virtuality $Q^2=\bm{k}^2-\omega^2 > 0$ requires a formation
time $t_{\rm coh}\sim \omega/Q^2$. ($\bm{k}$ is the parton 3--momentum,
and we assume high--energy kinematics: $|\bm{k}|\simeq \omega \gg Q$.)
This follows from the uncertainty principle: in a comoving frame where
the parton has zero momentum, its formation time is of order $1/Q$; this
becomes $\omega/Q^2$ after boosting by the parton Lorentz factor
$\gamma_p=\omega/Q$. Note also that, when the parent heavy quark is
highly energetic ($\gamma\gg 1$), the momentum $\bm{k}$ of the emitted
parton is predominantly longitudinal: $|\bm{k}|\simeq k_\ell \simeq
\omega$, whereas $k_\perp\sim Q\ll k_\ell$.

\texttt{(ii)} During the formation time $t_{\rm coh}\sim \omega/Q^2$, the
heavy quark does not radiate just a single parton, but rather a large
number of quanta, of $\order{\sqrt{\lambda}}$, whose emissions are
uncorrelated with each other. This is merely an assumption, which as we
shall see provides the right $\lambda$--dependence for the final results.

\texttt{(iii)} Only those quanta can be lost towards the plasma, whose
virtualities are small enough --- smaller than the {\em saturation
momentum} $Q_s\sim t_{\rm coh}T^2$ corresponding to the parton formation
time. This follows from the analysis in Ref.~\cite{Hatta:2007cs} which
shows that an energetic parton propagating through the strongly coupled
plasma feels the latter as a constant force $\sim T^2$ which acts towards
reducing its transverse momentum (or virtuality). Then, a space--like
parton, which would be stable in the vacuum, can decay inside the plasma
provided the lifetime $t_{\rm coh}$ of its virtual fluctuations is large
enough for the mechanical work $\sim t_{\rm coh}T^2$ done by the plasma
to compensate the energy deficit $\sim Q$ of the parton. This condition
amounts to $Q\lesssim Q_s$, with the upper limit given by
 \beq
 Q_s\,\sim\,t_{\rm coh}T^2\,\sim\,(\omega
 T^2)^{1/3}\,\sim\,\sqrt{\gamma_p}\, T\,,\eeq
where we have also used $t_{\rm coh}\sim \omega/Q^2$ and
$\gamma_p=\omega/Q$.

\texttt{(iv)} The rapidities of the radiated quanta are bounded by the
rapidity of the heavy quark: $\gamma_p\lesssim\gamma$. This is again
motivated by the uncertainty principle and at least at weak coupling it
is confirmed by the explicit construction of the heavy quark wavefunction
\cite{Dokshitzer:2001zm}.

We shall now use this picture to compute the rate for energy loss and
momentum broadening of the heavy quark. The latter radiates energy
$\Delta E\sim \sqrt{\lambda}\omega$ over a time interval $\Delta t\sim
\omega/Q^2$, where $\omega$ and $Q$ are constrained by $Q\lesssim
Q_s(\omega,T)$. The dominant contribution to the rate $|\Delta E/\Delta
t|$ comes from those quanta carrying the maximal possible energy
$\omega\simeq \gamma Q$ and also the maximal corresponding virtuality
$Q\simeq Q_s(\gamma,T) \sim \sqrt{\gamma}\, T$ (to minimize the emission
time). Therefore,
 \beq\label{dEdt}
 -\,\frac{\rmd E}{\rmd t}\,\simeq\,
 \frac{\sqrt{\lambda}\, \omega}{(\omega/Q_s^2)}
 \,\simeq\,\sqrt{\lambda}\,Q_s^2 \,\sim\,\sqrt{\lambda}\,\gamma\,T^2
 \,,\eeq
in qualitative agreement with the estimate for the drag force $F_{\rm
drag}=\eta\gamma v\sim  \gamma\sqrt{\lambda}T^2$ in \eqnum{LangPL}.
(Recall that we consider the relativistic case $v\simeq 1$.)

Consider similarly momentum broadening: being uncorrelated with each
other, the $\sqrt{\lambda}$ quanta emitted during a time interval $t_{\rm
coh}$ have transverse momenta which are randomly oriented, so their
emission cannot change the {\em average} transverse momentum of the heavy
quark. However, the changes in the {\em squared} momentum add
incoherently with each other, thus yielding (once again, the dominant
contribution comes from quanta with $Q\sim Q_s(\gamma,T)$ and
$\omega\simeq \gamma Q$)
 \beq\label{dpTdt}
 \frac{\rmd \langle p_\perp^2\rangle}{\rmd t}\,\sim\,
\frac{\sqrt{\lambda}\,Q_s^2}{(\omega/Q_s^2)} \,\sim\,
\sqrt{\lambda}\,\frac{Q_s^4}{\gamma Q_s}\,\sim\,
 \sqrt{\lambda}\,\sqrt{\gamma}\,T^3\,,\eeq
which is parametrically the same as the estimate for $\kappa_\perp$ in
the first equation \eqref{kappa}. The random emissions also introduce
fluctuations in the energy (or longitudinal momentum) of the heavy quark,
in addition to the average energy loss. The dispersion associated with
such fluctuations is estimated similarly to \eqnum{dpTdt} (below, $\delta
p_\ell \equiv p_\ell - \langle p_\ell\rangle$)
 \beq\label{dpLdt}
 \frac{\rmd \langle \delta p_\ell^2\rangle}{\rmd t}\,\sim\,
\frac{\sqrt{\lambda}\,\omega^2}{(\omega/Q_s^2)} \,\sim\,
\sqrt{\lambda}\,\sqrt{\gamma}\,\gamma^2\,T^3\,,
 \eeq
in qualitative agreement with the previous result, \eqnum{kappa}, for
$\kappa_\ell$. Note that, with this interpretation, the relative factor
$\gamma^2$ in between $\kappa_\ell$ and $\kappa_\perp$ is simply the
consequence of the relation $\omega\simeq \gamma Q$ between the energy
and the virtuality (or transverse momentum) of an emitted parton.

This physical picture also clarifies the role of the world--sheet horizon
in the dual gravity calculation: via the UV/IR correspondence, the radial
position $z_s=1/\sqrt{\gamma}$ of this horizon (in units of $z_H=1/\pi
T$) is mapped onto the saturation momentum $Q_s \sim \sqrt{\gamma}\, T$
in the boundary, gauge, theory. Hence the emergence of the noise terms
from the near--horizon dynamics of the string reflects
quantum--mechanical fluctuations in the emission of quanta with
virtualities $Q\sim Q_s$, which as we have just seen control momentum
broadening.

\begin{figure}
\centerline{
\includegraphics[width=0.65\textwidth]{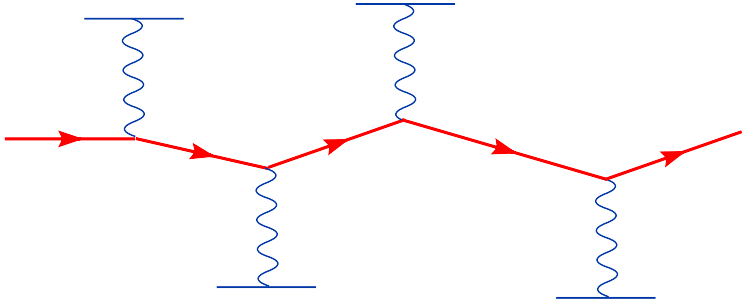}
}
\caption{\sl Momentum broadening via thermal rescattering at weak coupling.
\label{fig:BroadpQCD}}\end{figure}

\begin{figure}
\centerline{
\includegraphics[width=0.65\textwidth]{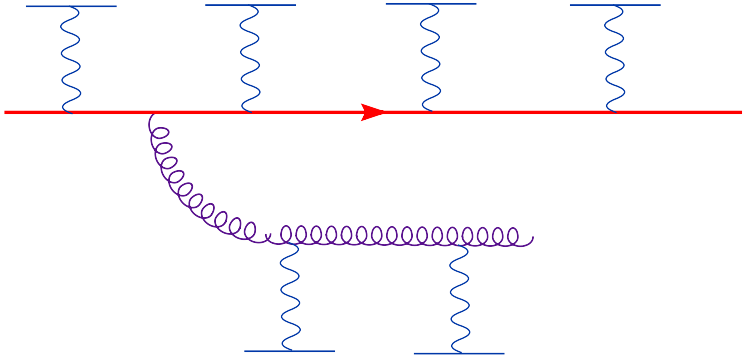}
}
\caption{\sl Energy loss via medium--induced
gluon emission at weak coupling\label{fig:EnlosspQCD}}\end{figure}

It is furthermore interesting to compare the above physical picture to
the corresponding one at weak coupling
\cite{BDMPS,Baier:2002tc,Kovner:2003zj,CasalderreySolana:2007zz}. Note
first that the mechanism for momentum broadening is different in the two
cases: at weak coupling, this is dominated by thermal rescattering, i.e.,
by successive collisions with the plasma constituents which are thermally
distributed (see Fig.~\ref{fig:BroadpQCD}). In that case, the rate ${\rmd
\langle p_\perp^2\rangle}/{\rmd t}\equiv\hat q$ defines a genuine
transport coefficient --- the ``jet--quenching parameter'' ---, i.e. a
local quantity which depends only upon the local density of thermal
constituents (quarks and gluons) together with the gluon distribution
produced via their high--energy evolution. By contrast, at strong
coupling, the dominant mechanism at work is medium--induced radiation,
which is intrinsically non--local (it requires the formation time $t_{\rm
coh}$) and hence cannot be expressed in terms of a local transport
coefficient. Medium--induced radiation is of course possible at weak
coupling too (see Fig.~\ref{fig:EnlosspQCD}), but the respective
contribution is suppressed by a factor $g^2N_c$ as compared to the
thermal rescattering. We see that, formally, it is the replacement
$g^2N_c\to \sqrt{\lambda}$ (i.e., the coherent emission of a large number
of quanta) which makes the medium--induced radiation become the dominant
mechanism for momentum broadening at strong coupling.

On the other hand, energy loss is predominantly due to medium--induced
radiation at both weak and strong coupling, but important differences
occur between the detailed mechanisms in the two cases (compare
Figs.~\ref{fig:Broad} and \ref{fig:EnlosspQCD}): At weak coupling, the
radiated gluon, which typically comes from a highly virtual gluon in the
quark wavefunction, is freed (radiated) via thermal rescattering. At
strong coupling, radiation is caused by the plasma force $\sim T^2$.
After being emitted, the parton undergoes successive medium--induced
branchings, thus producing a system of partons with lower and lower
energies and transverse momenta, down to values of $\order{T}$, when the
partons cannot be distinguished anymore from the thermal bath.

It is finally interesting to notice that, in spite of such physical
dissymmetry, the formula for energy loss at weak coupling can be written
in a form which ressembles \eqnum{dEdt}, namely
 \beq\label{dEdtw}
 -\,\frac{\rmd E}{\rmd t}\,\simeq\,g^2N_c\,Q_s^2\qquad
 \mbox{(weak coupling)}\,,\eeq
where however $Q_s$ is now the saturation momentum to lowest order in
perturbative QCD and is related to the respective jet--quenching
parameter via $Q_s^2\simeq \hat q t_{\rm coh}$. Energy loss involves a
coherent phenomenon at both weak and strong coupling.

\bigskip

\bigskip

\subsection*{Acknowledgments}

We would like to thank Iosif Bena for useful discussions and Derek Teaney
for explaining some aspects of his recent work to us. G.C.~G. is
supported by Contrat de Formation par la Recherche, INSTN, CEA--Saclay.
The work of E.~I. is supported in part by Agence Nationale de la
Recherche via the programme ANR-06-BLAN-0285-01. The work of A.H.~M. is
supported in part by the US Department of Energy.


\providecommand{\href}[2]{#2}\begingroup\raggedright\endgroup

\end{document}